\mag=\magstephalf
\pageno=1
\input amstex
\documentstyle{amsppt}
\TagsOnRight
\interlinepenalty=1000
\NoRunningHeads

\pagewidth{16.5 truecm}
\pageheight{23.0 truecm}
\vcorrection{-1.0cm}
\hcorrection{-1.2cm}
\nologo

\NoBlackBoxes

\define \ee{\roman e}

\define \ordp{{\roman{ord}_p}}
\define \ult{\beta}
\define \ordu{{\roman{ord}_{\beta}}}
\define \QQ{{\Bbb Q}}
\define \RR{{\Bbb R}}
\define \ZZ{{\Bbb Z}}

{\centerline{\bf{ $p$-adic Difference-Difference
 Lotka-Volterra Equation and Ultra-Discrete Limit}}}

\author
\endauthor
\affil
Shigeki MATSUTANI\\
8-21-1 Higashi-Linkan Sagamihara, 228-0811, Japan \\
\endaffil
\endtopmatter


\subheading{Abstract}
 In this article, we have studied the 
 difference-difference Lotka-Volterra equations in
 $p$-adic number space and its $p$-adic valuation
 version. We pointed out that the structure of the
 space given by taking the ultra-discrete limit is
 the same as that of the $p$-adic valuation space. 


\document

\vskip 1 cm
{\centerline{\bf{Introduction}}}
\vskip 0.5 cm

In soliton theory, difference-difference equations, whose domain 
space-time are given by integers, and the ultra-discrete 
difference-difference equations, whose, all,  domain and range are given
by integers, are currently studied  [HT, TS, TTMS]. 

On the other hand, recently number theory and physics might be 
considered as a missing link of each other. For example, a set of 
geodesics in a compact Riemannian surface with genus $g\ge 2$ are 
investigated in the framework of {\it chaos} because any geodesics, or
orbits, part from each other due to its negative curvature [V,Su] 
(whereas the Jacobi varieties of the Riemannian surfaces are completely
classified by a {\it soliton} equation [M]). By quantization of the 
orbits, there appears {\it quantum chaos} and, as it is very mysterious,
its partition function has very resemble structure of zeta functions in
number theory [V,Su]. (Level statistics in {\it quantum choas} is also
connected with the integrable system [So].)  Using the resemblance of
zeta functions, Connes proposed a kind of unification of number theory
and quantum statistical physics in order to solve the Riemannian 
conjecture of $\zeta$-function [BC,Co].

Further on the discrimination problem of integrability of Hamiltonian
system, there appears Galois theory in the category of differential 
equation [MR, Y], which plays the same role in the category of the 
number theory.

Thus in order to know what is the integrability or quantization,
it is not surprising that there appears integer theory in physics.
In fact, there are many other studies
pointing out that the $p$-adic number theory and
non-archimedean valuation theory are
closely related to statistical and quantum physics
[RTV, BF, VVZ], even though $p$-adic space has a metric which 
differs from euclieadean sense. 
These correspondences might imply that there is a deep hidden symmetry
behind physics and number theory and give a novel step
to mathematical physics.

Thus I believe that it is very important to 
interpret  such recent development of soliton 
theory  using $p$-adic number theory and
non-archimedean valuation theory.

In this article, we will mainly deal with the Lotka-Volterra equation
as a typical difference-difference soliton equation.
We will show that 1) even in $p$-adic
space of the number theory, 
the $p$-adic difference-difference Lotka-Volterra equation has mathematical
meanings and has solutions.  Next we will define  
the $p$-adic valuation version of the equation. Then we will show that
2)  a quantity obtained by the ultra-discrete limit in the soliton theory
should be regarded as the non-archimedean valuation
as $p$-adic space has the $p$-adic valuation  [C,VVZ].

In this article we will start from a
preliminary of $p$-adic number theory.
Next we
will review  the recent development of difference-difference 
and the ultra-discrete soliton theory [TTMS]. We will deal with the 
difference-difference and ultra-discrete difference-difference
Lotka-Volterra equations. 
After we  formally construct a 
$p$-adic difference-difference Lotka-Volterra equation, we
will investigate  its existence and explicit forms of its solutions. 
Then we will show that even in the $p$-adic space
the difference-difference Lotka-Volterra equation has mathematical
meanings and has soliton solutions.
Next by computations of $p$-adic valuation,
we will show resemblance
between the $p$-adic valuation of the $p$-adic 
difference-difference Lotka-Volterra equation and  ultra-discrete 
difference-difference Lotka-Volterra equation. 
Then it will be shown that the ultra
discrete limit has the same structure as the 
$p$-adic valuation. Finally we will comment upon 
physical and mathematical meanings of the 
ultra-discrete limit.

\vskip 0.5 cm

{\centerline{\bf{ Preliminary: $p$-adic Space}}}

\vskip 0.5 cm

Let us consider $p$-adic field $\QQ_p$ for a prime
number $p$ [BF. C, I, RTV, VVZ]. For a rational number $u \in
\QQ$ which are given by $u = \dfrac{v}{w} p^m$ ($v$
and $w$ are coprime to the prime number $p$ and $m$
is an integer), we will define a symbol $[[u]]_p = p^m$.
Here we will define  the $p$-adic 
valuation of $u$ as a map from $\QQ$ to a set of 
integers $\Bbb Z$, 
$$
 \ordp(u):=  \log_p [[u]]_p, \ \text{ for } u \neq 0 , 
 \ \text{ and }
	\ \ordp(u):= \infty,\  \text{ for } u = 0. 
$$
This valuation has following properties (I${}_p$); 

\subheading{I${}_p$:}
For $u,v \in \QQ$,
\roster

\item $ \ordp(u v ) = \ordp(u) + \ordp( v ) $. 

\item $ \ordp(u + v ) \ge \min( \ordp(u) , \ordp(
v ))$. 
 
 If $\ordp(u) \neq \ordp(v)$, 
$ \ordp(u + v ) = \min( \ordp(u) , \ordp( v ))$.

\endroster

\vskip 0.5 cm

This property (I${}_p$-1) means that $\ordp$ is a 
homomorphism from the multiplicative group 
$\QQ^\times$ of $\QQ$ to the additive group $\ZZ$.
The $p$-adic metric is given by $|v|_p = 
p^{-\ordp(v)}$. It is obvious that it is a metric
because it has the properties (II${}_p$); 

\subheading{II${}_p$:}
For $u,v \in \QQ$,
\roster

\item if $ |v|_p =0 $, $v$=0.

\item $ |v|_p \ge0 $.

\item $ |v u |_p= |v|_p |u|_p $.

\item $|u + v|_p \le\max(|u|_p, | v|_p) 
\le |u|_p+|v|_p$.

\endroster

\vskip 0.5 cm

The $p$-adic field $\QQ_p$ is given as a completion
of $\QQ$ with respect to this metric so that
properties (I${}_p$) and (II${}_p$) survive for $\QQ_p$. 

Further, we note  that $\ZZ_p$, integer part of $\QQ_p$,
is a "localized ring" and has only prime ideals $\{0\}$
and $p \ZZ_p$.
As the properties of $p$-adic metric, the 
convergence condition of series $ \sum_{m} x_m$ is identified with
the vanishing condition of sequence $|x_m|_p \to 0$ for $m \to \infty$
due to the relationship,
$$
      | \sum_{m} x_m |_{p} = 
	\max |x_m|_p.  \tag 1
$$


\vskip 0.5 cm

Let us define $|u|_\infty$ as a natural metric or absolute value over
real field $\Bbb R$, $|u|_\infty := |u|$, and  $\Bbb
R$ is regarded as $\infty$ point of prime numbers;
we will denote $\Bbb R$ as $\QQ_\infty$. Then we 
have a relation for any non-zero $u \in \QQ$,
$$
	\prod_{p \in \frak A} |u|_p = 1,
$$
where $ \frak A$ is a set of prime numbers and 
$\infty$. This is an adelic property of $p$-adic metric.

\vskip 0.5 cm

{\centerline{\bf{  Difference-Difference Lotka-Volterra Equation}}}

\vskip 0.5 cm

Here we will review the difference-difference Lotka-Volterra equation
[HT, TTMZ]. First, we will consider the Korteweg-de Vries (KdV) equation,
$$
	\partial_t u + 6 u \partial_s u + 
	\partial_s^3 u = 0, \tag 2 
$$
where $\partial_t := \partial /\partial t$ and 
$\partial_s := \partial /\partial s$ and $u =u 
(s,t)$ whose domain $(t,s)$ is $\RR^2$. This 
differential equation was found by Korteweg and de
Vries. However this differential equation and 
related (KdV) hierarchy were also discovered by 
Baker about one hundred years ago [B]. He studied
the hyperelliptic functions and essentially 
discovered KdV hierarchy as differential equations
generating periodic and algebraic functions over 
a hyperelliptic curve. He defined  hyperelliptic
$\sigma$ function as the best tuning theta 
function there and hyperelliptic $\wp$ functions as
meromorphic functions over the hyperelliptic curve;
{\it e.g.}, they are connected as $\wp= \partial_s^2 \log 
\sigma$. To evaluate the $\wp$-functions, he also 
used Paffians a bilinear operator, 
 and bilinear equations, which, latter two, are recently 
called Hitora bilinear operator and bilinear equations [B].
These $\sigma$ and $\wp$ functions are closely
related to the problems in number theory [O].

Along the line of arguements of [TTMS], we will deal with
the difference-difference Lotka-Volterra as a 
difference-difference analogue of (2) here and
next ultra-discrete difference-difference Lotka-Volterra.

The difference-difference Lotka-Volterra
equation is given as [HT], 
$$
 \frac{c^{m+1}_n}{ c^m_n} = \frac{ 1 + \delta 
 c^m_{n-1}} { 1 + \delta c^{m+1}_{n+1} }.
 \tag 3
$$
According to the arguments in [DJM1-3, H, HT, TTMS], 
(3) is related to the bilinear difference-difference
equation,
$$
\tau^{m+1}_{n+1} 
\tau^{m}_n-(1+\delta)\tau^{m}_{n+1}\tau^{m+1}_{n}
+ \delta \tau^m_{n-1} \tau^{m+1}_{n+2} = 0, \tag 4
$$
where
$$
	c_n^m=\frac{ \tau^m_{n-1}\tau^{m+1}_{n+2}}
	{\tau^{m}_{n}\tau^{m+1}_{n+1}}.
	\tag 5
$$ 
For example, the two-soliton solution is expressed as [H],
$$
	\tau_n^{m}= 1+ \ee^{\eta_1(m,n)} 
	+\ee^{\eta_2(m,n)} + A \ee^{\eta_1(m,n)+ 
	\eta_2(m,n)}, \tag 6
$$
where $k_a$, $\omega_a$ 
$\eta^0_a$ ($a=1,2$) are real numbers satisfied with,
$$
\split
\eta_a(m,n) &= k_a n - \omega_a m  +\eta^0_a,\\
\ee^{\omega_a} &=
\frac{ 1+ \delta ( \ee^{k_a}+1)}
{ 1+ \delta ( \ee^{-k_a}+1)}, \\
A&= \frac{\sinh^2[( k_1- k_2)/2]}{\sinh^2[( k_1+ k_2)/2]}.
\endsplit \tag 7
$$

Similarly, we have more general solutions [H, 
DJM1-3, TTMS].

\vskip 0.5 cm

{\centerline{\bf{ Ultra-Discrete Space}}}

\vskip 0.5 cm

Next we will introduce the ultra-discrete limit 
following [TTMS]. In order to make our argument easy,
we will change its notation but its essential
definition does not differ.

 Let $\Cal A_{\beta}$ be a subset
of non-negative functions over $(m,n,\beta) \in \Bbb
Z^2 \times (\RR_{>0})$  where $\RR_{>0}$ is a set of
positive real numbers. 
Here we regard that $\beta$ is a parameter of the function
over $\ZZ^2$. 

Let us define a map 
$\ordu:\Cal A_{\beta} \to \RR$. We set $ \ordu(0
) = \infty$  for zero and for $u \in \Cal A_{\beta}$,  
$$
	\ordu(u):= 
	-
	\lim_{ \beta \to +\infty} \frac{1}{\beta}\log ( u). \tag 8
$$ 
$\Cal A_{\beta}$ is characterized so that $\ordu(u)$ has a meaning value.

Typically it behaves like, 
$$
	\ordu(\ee^{-\beta A } + \ee^{-\beta B }
	+ \cdots)
	= \min ( A, B, \cdots) .
$$
We note that  this map $\ordu$ has the properties (I${}_\beta$); 

\subheading{ I${}_\beta$}

For
$u, v \in \Cal A_{\beta}$,
\roster

\item $ \ordu(u v ) = \ordu(u) + \ordu( v ) $.

\item $ \ordu(u + v ) =\min( \ordu(u) , \ordu( v ))$.

\vskip 0.5 cm

\endroster

We note that this is a non-archimedean valuation 
because for $A > B$, there does not exist a finite
integer $n$ such that $\ordu(\ee^{-\beta A })< 
\ordu(n \ee^{-\beta B })$ [I,C, VVZ].
First we will note that
this valuation is resemble to the property I${}_p$ of $p$-adic valuation.
We will call it ultra-valuation.

By introducing new variables $f^m_n := - \ordu( 
c^m_n )$ and $d := - \ordu( \delta )$ [T],
we have a ultra-valuation version of the difference-difference 
Lotka-Volterra equation (3),
$$
f^{m+1}_n - f^m_n = \ordu( 1 + \delta_p c^m_{n-1})
- \ordu(1 + \delta_p c^m_{n-1})\tag 9
$$
or
$$
f^{m+1}_n - f^m_n = \max( 0 , f^m_{n-1} + d) - \max(
0,  f^{m+1}_{n+1} + d). \tag 9'
$$

This is a ultra-discrete difference-difference  
Lotka-Volterra equation, which is also integrable;
its integrability was proved in [TTMS].
Of course, when $f$'s are given by quantities of integers
times $d$ respectively, we can normalize it as $d=1$ by dividing $d$.

\vskip 0.5 cm

{\centerline{\bf{$p$-adic  Difference-Difference 
 Lotka-Volterra Equation and Its Valuation Version }}}

\vskip 0.5 cm

Next we will show that even in the 
$p$-adic space, difference-difference  Lotka-Volterra equation
has mathematical meaning and has solutions.

First we will formally introduce the $p$-adic  
difference-difference  Lotka-Volterra equation for
a $p$-adic series $\{c_n^m\in \QQ_p\}$ ($p \neq 2$),
$$
 \frac{c^{m+1}_n}{ c^m_n} = \frac{ 1 + \delta_p 
 c^m_{n-1}} { 1 + \delta_p c^{m+1}_{n+1} },
 \tag 10
$$
where $ \delta_p \in p\ZZ_p$. Noting that from (1),
$ p\ZZ_p$ is domain of exponential function  and 
$1+p\ZZ_p$ is domain of logarithm function [VVZ].
Further addition of elements of $p\ZZ_p$ belongs
to $p\ZZ_p$ because $p\ZZ_p$ is an ideal.
If $k_a$ and $\eta_a^0$ ($a=1,2)$ belong to $p\ZZ_p$,  two-soliton
solution (6) and the conditions (7) replacing $\delta$ with $\delta_p$
 are well-defined in $p$-adic space. In fact, in the equation,
$$
	\omega_a= \log \frac{ 1+ \delta_p ( \ee^{k_a}+1)}
{ 1+ \delta_p ( \ee^{-k_a}+1)}
$$
$(1+ \delta_p ( \ee^{k_a}+1))/( 1+ \delta_p ( \ee^{-k_a}+1))$ 
can be expanded in $p$-adic space and belongs to $1+p\ZZ_p$
and $\omega_a$ has a value in $\QQ_p$. 
Since region of logarithm function for $1+p\ZZ_p$ is $p\ZZ_p$,
$\omega_p$ is an element of $p\ZZ_p$. Due to the properties of
ideal, $\eta_a(m,n) := k_a n - \omega_a m  +\eta^0_a$ is
in a domain of exponential function  in $p$-adic space.
Further $p$-adic version of 
$A$ in (7) also can be computed. Hence $p$-adic
version $\tau$ in (6) and $c$ has  a finite value in $p$-adic space.
In other words, one- and two-soliton
solutions exist in (10).

For the case of $p = 2$, since $ 4\ZZ_2$ is domain of exponential function,
$ \delta_2$, $k_a$ and $\eta_a^0$ ($a=1,2)$ belong to $4\ZZ_2$.
Further though $k_a$ must also be satisfied
with $k_1 \pm k_2 \in 8 \ZZ_2$, we can argue it in
similar way.

Further similarly we can construct other soliton solutions for 
$p$-adic equation (10) following
the procedure in [H, DJM1-3, TTMS].

Hence it is very surprising that
(10) does not has only formal meaning but
also mathematical meanings in the $p$-adic
space. In other words, they has $n$-soliton solutions.
We emphasize that
the fact that even in $p$-adic space,
the difference-difference equation has a soliton solution
 stands upon a very subtle balance and  is
absolutely different from trivial phenomena.

Due to the properties of $p$-adic space, there are
positive integers $n$ and $m$ such that these
solutions module $p^n$  are
satisfied with (10) module $p^m$.
For example, $p\gg 1$ and $n=3$ case:
$\omega_a \equiv 2\delta_p k_a - 4\delta_p^2$ module $p^3$
and $A\equiv\dfrac{k_1-k_2}{k_1+k_2} 
\left(1 + \dfrac{1}{24}(k_1^2+k_2^2)\right)$
module $p^3$. Since $\exp(p)\equiv 1+ p + \frac{1}{2} p^2$ modulo $p^3$,
$\tau$'s and $c$'s are determined in $p^3$. 

As the $p$-adic difference-difference Lotka-Volterra equation
is well-defined, we will consider the $p$-adic valuation
of the equation.
By letting $f^m_n := - \ordp( c^m_n )$ and 
$d_p:=-\ordp( \delta_p)$, we have
$$
f^{m+1}_n - f^m_n = \ordp( 1 + \delta_p c^m_{n-1})
- \ordp(1 + \delta_p c^m_{n-1}).
\tag 11
$$
When we assume that $f^{m}_{n}\neq -d_p$, (11) becomes
$$
f^{m+1}_n - f^m_n = \max( 0 , f^m_{n-1} + d_p) - 
\max( 0,  f^{m+1}_{n+1} + d_p).
\tag 12
$$
It is also  surprising that (12) has the same form
as the ultra-discrete difference-difference Lotka-Volterra 
equation (9) at all. 

In other words, we conclude that the structure
of ultra-discrete limit has the same as that in 
$p$-adic analysis.

\vskip 0.5 cm

{\centerline{\bf{ Ultra-Discrete Metric From Point 
of View of Valuation Theory}}}

\vskip 0.5 cm

As we saw the similarity between ultra valuation and
$p$-adic valuation, we will construct the ultra 
metric following the definition of $p$-adic metric.

Since soliton theory is defined over the field
whose characteristic is zero, 
we might regard it  as theory of
$\QQ_\infty$.
However it should be also noted that since the 
ultra-valuation is a natural non-archimedean
valuation of real valued functions,
another real valued metric is naturally defined, 
which seem to slightly differ from the 
ordinary metric $|x|_\infty\equiv |x|$.

By
introducing a real number   $\bar \beta \gg 1$, 
it is defined as
$$
	|x|_{\ult} := \left
	(\ee^{ -\bar\beta}\right)^{\ordu(x)},
$$
which is a kind of exponential valuation [I].
We call this ultra-metric. It has properties (II${}_\beta$); 

\subheading{II${}_\beta$}

\roster

For $u$, $v \in \Cal A_{\beta}$, ($|v|_{\ult}$ depends upon $\bar \beta$)

\item if $ |v|_{\ult} =0 $, $v$=0.

\item $ |v|_{\ult} \ge0 $.

\item $ |v u |_{\ult}= |v|_{\ult} |u|_{\ult} $.

\item $|u + v|_{\ult} \le |u|_{\ult}+|v|_{\ult}$.

\endroster
\vskip 0.5 cm

Since the ultra-discrete and the $p$-adic valuation are given by 
for $u \in \Cal A_\beta$ and $v\in \QQ_p$, ($u\neq 0$, $v\neq0$),
$$
\ordu(u)=\lim_{ \beta \to +\infty} \log_{\ee^{-\beta}} ( u),
\quad \text{and}\quad
  \ordp(v)=\log_p [[v]]_p,
$$
$\ee^{-\beta}|_{\beta \to \infty}$ plays the same role of $p$.
However it should be noted that since this ultra valuation
is defined in $\RR$, $|x|_{\ult}$  is defined
above rather than $\left(\ee^{-\bar \beta}\right)^{-\ordu(x)}$ 
whereas $|x|_p = p^{-\ordp x}$.
As we have assumed that $x \in \Cal A_{\beta}$ has 
a value at $\beta\to\infty$,
$$
\left.{|x|_{\ult}}\right|_{\bar \beta\sim \infty} \sim  
\left.\exp( - \bar \beta (-(\log x)/\beta))\right|
_{\bar \beta \sim \beta \sim \infty}
=|x|^{  \bar \beta/\beta}|_{\bar \beta \sim \beta \sim \infty},
$$
and thus, as it is not guaranteed, it may be regarded that $|x|_{\ult}
\sim |x|$, in heart, by synchronizing $\bar \beta$ and $\beta$. It 
implies that the ultra-metric $|x|_\ult$ might be also considered as the
natural metric at $\QQ_\infty$.

In this metric, the convergence condition of series
is also equivalent with the vanishing condition
of sequence and we have the relation, 
$$
	| \sum_m x_m |_{\ult} = \ee^{-\bar \beta 
	\min(\ordu(x_m)) }. \tag 13
$$
We should note that this metric appears in the low
temperature treatment of the statistical physics and
in the semi-classical treatment of path integral 
[D, FH]. For the low temperature limit $\bar \beta
\sim\beta=1/T$, $T \to 0$ or the  classical limit
of deformation parameter $\bar \beta \sim 
\beta=1/\hbar$, $\hbar\to 0$, only the minimal point
survives and contributes to zero temperature or 
classical phenomena. Thus the ultra-discrete limit
is sometimes called "quantization", as
a terminology of  discretaization in digital picture, 
in the literature
in the soliton theory but it should be regarded as
low temperature limit of statistical mechanical 
phenomena or classical limit of quantum phenomena.
The reason why the domain of $\Cal A_{\beta}$ is
non-negative might be related to the positiveness
of the probability. 

Then there  arises a question why the ultra-discrete
limit is related to integer valued solutions for a
soliton solution. Function form of finite type 
solution of (1) including soliton solution is 
completely determined at the infinity point of the
spectral parameters $k=\infty$ [DJM1, K]. The 
soliton solution is given by exponential function
whose power is polynomial of $(k,s,t)$ owing to 
algebraic properties of soliton solutions. Since 
polynomial of integer valued $(k,s,t)$ is also 
integer, ultra-discrete is associated with integer
valued solutions.

Further it is known that some
of properties in the $q$ analysis can be regarded
as those in $p$-adic analysis by setting $q=1/p$ 
[VVZ]. We have correspondence among $p$, $q$ and $
\ee^{\beta}$ as
$$
\ee^{- \beta}  \Longleftrightarrow p
 \ (\beta \sim \infty), 
\quad p \Longleftrightarrow 1/q, \quad
q \Longleftrightarrow  \ee^{\beta}\ (\beta \sim 0).
$$
From quantum mechanical point of view, it must be
emphasized that the classical regime appears as a
non-archimedean valuation, which is an algebraic 
manipulation, for quantum mechanical values. 
In this analogy, we might regard that 
$\ZZ$ is in a 
classical regime whereas $\QQ_p$'s $(p \in \frak A)$
are of quantum world in number theory.
In fact, Bost and Connes applied the method in quantum statistical
field theory (or methods in type III factor of non-commutative
ring) to the problem of number theory
and proposed a strategy to Riemannian hypothesis of 
$\zeta$-function [BC,Co].

 Even though the non-degenerated
 hyperelliptic functions might not be related to the
 discrete Lotka-Volterra equation (3), the discrete
 Lotka-Volterra equation are related to invariant
 theory as their conserved quantities [HT]. Thus 
 this $p$-adic approach might give another aspect
 on relations between soliton and number theory besides
 [IMO, O, P].
 Further as mentioned in the introduction,
 $p$-adic analysis is closely connected with quantum and
 statistical physics [BC, Co, BF, RTV, VVZ].
 In fact, due to their translation symmetry, 
 the Green functions (or the correlation functions)
in the quantum and statistical physics usually
have the properties of Toeplitz matrix and their product are
given by a convolution.
A difference-difference  soliton equation, in general, can be
regarded as an identity of Toeplitz determinant [So].
On the other hand, the  Heck algebra in number theory
is defined using a convolution in the adelic space.
Thus it is natural to expact that they should be
written in single framework.
Accordingly we hope that the correspondence between 
 $p$-adic  and ultra-discrete structures 
 might have an effect on these and other studies
 on physics.

\vskip 0.5 cm

{\centerline{\bf{ Acknowledgment}}}

\vskip 0.5 cm

I would like to thank Prof. T. Tokihiro, Prof.
Y. \^Onishi and members of Toda seminar.

\enddocument